\documentclass[a4paper]{article}
\usepackage[utf8]{inputenc}
\usepackage{amsmath, amsfonts, amsthm, amssymb, bbm, bm}
\usepackage{mathrsfs}
\usepackage{centernot}
\usepackage{enumerate}
\usepackage[shortlabels]{enumitem}
\usepackage{tabularx}
\usepackage{multicol}
\usepackage{multirow}
\usepackage[colorlinks=true,linktoc=all, allcolors=black]{hyperref}
\usepackage{setspace}
\usepackage[margin=1in]{geometry}
\usepackage[round]{natbib}
\usepackage{booktabs}
\usepackage{tikz}
\usetikzlibrary{shapes,arrows,calc}
\usepackage{algorithm} 
\usepackage{algpseudocode} 
\usepackage{authblk}
\usepackage[english]{babel} 
\usepackage{blindtext}

\newcommand{\E}{\mathbb{E}}
\newcommand{\D}{\mathcal{D}}

\DeclareMathOperator{\expit}{expit}
\DeclareMathOperator{\logit}{logit}

\newcommand{\X}{\mathbb{X}}

\DeclareMathOperator{\Bernoulli}{Bernoulli}

\DeclareMathOperator{\var}{Var}

\newcommand{\cmid}{\,|\,}

\newcommand\indep{\protect\mathpalette{\protect\independenT}{\perp}}
\def\independenT#1#2{\mathrel{\rlap{$#1#2$}\mkern2mu{#1#2}}}

\theoremstyle{plain}
\newtheorem{lem}{Lemma}[section]
\newtheorem{thm}{Theorem}

\theoremstyle{definition}

\newtheorem{rmk}{Remark}
\newtheorem{exm}{Example}

\newcommand{\bitem}{\begin{itemize}}
\newcommand{\eitem}{\end{itemize}}

\newcommand{\barr}{\begin{array}}
\newcommand{\earr}{\end{array}}

\newcommand{\bmat}{\begin{pmatrix}}
\newcommand{\emat}{\end{pmatrix}}

\newcommand{\bs}{\boldsymbol}

\def\bal#1\eal{\begin{align*}#1\end{align*}}

\newtheorem{assumption}{Assumption}

\newcolumntype{L}{>{$}l<{$}}
\newcolumntype{C}{>{$}c<{$}}

\usepackage{tikz}
\usetikzlibrary{shapes, arrows, arrows.meta, calc, positioning}

\tikzset{nv/.style={circle, color=red, fill=red, inner sep=0.5mm}}
\tikzset{rv/.style={circle, draw, thick, minimum size=7mm, inner sep=0.5mm}}
\tikzset{fv/.style={rectangle, draw, thick, minimum size=7mm, inner sep=0.5mm}}
\tikzset{lv/.style={circle, color=red, fill=gray!30, draw, thick, minimum size=7mm, inner sep=0.5mm}}
\tikzset{rve/.style={ellipse, draw, thick, minimum size=7mm, inner sep=0.5mm}}

\tikzset{rvs/.style={circle, draw, thick, minimum size=6mm, inner sep=0.5mm}}
\tikzset{fvs/.style={rectangle, draw, thick, minimum size=6mm, inner sep=0.5mm}}
\tikzset{lvs/.style={circle, color=red, fill=gray!30, draw, thick, minimum size=6mm, inner sep=0.5mm}}
\tikzset{rves/.style={ellipse, draw, thick, minimum size=6mm, inner sep=0.5mm}}

\tikzset{deg/.style={->, very thick, color=blue}}
\tikzset{degl/.style={->, very thick, color=red}}
\tikzset{beg/.style={<->, very thick, color=red}}
\tikzset{cdeg/.style={{Circle[length=+2pt 2.5,width=+2pt 2.5, fill=none]}->, very thick, color=blue}}
\tikzset{cceg/.style={{Circle[length=+2pt 2.5,width=+2pt 2.5, fill=none]}-{Circle[length=+2pt 2.5,width=+2pt 2.5, fill=none]}, very thick}}
\tikzset{uceg/.style={{Circle[length=+2pt 2.5,width=+2pt 2.5, fill=none]}-, very thick}}
\tikzset{ueg/.style={very thick}}

\definecolor{oxblue}{RGB}{0, 33, 71}

\title{Combining experimental and observational data through a power likelihood}

\author[1]{Xi Lin}
\author[2]{Jens Magelund Tarp}
\author[1]{Robin J. Evans}
\affil[1]{Department of Statistics, University of Oxford, UK}
\affil[2]{Novo Nordisk, Denmark}
\date{\today}

\begin{document}

\maketitle



\begin{abstract}
Randomized controlled trials are the gold standard for causal inference and play a pivotal role in modern evidence-based medicine. However, the sample sizes they use are often too limited to draw significant causal conclusions for subgroups that are less prevalent in the population. In contrast, observational data are becoming increasingly accessible in large volumes but can be subject to bias as a result of hidden confounding. Given these complementary features, we propose a power likelihood approach to augmenting RCTs with observational data to improve the efficiency of treatment effect estimation. We provide a data-adaptive procedure for maximizing the expected log predictive density (ELPD) to select the learning rate that best regulates the information from the observational data. We validate our method through a simulation study that shows increased power while maintaining an approximate nominal coverage rate. Finally, we apply our method in a real-world data fusion study augmenting the PIONEER 6 clinical trial with a US health claims dataset, demonstrating the effectiveness of our method and providing detailed guidance on how to address practical considerations in its application.


\end{abstract}

\newpage

\section{Introduction}\label{sec:intro}


Experimental data and observational data represent two distinct regimes for causal inference. Experimental data are collected through designed experiments with randomized interventions; a typical example is randomized controlled trials (RCTs). Randomized intervention isolates the causal effect of interest from unwanted and potentially unobserved confounding factors, as the experimental protocol ensures---in expectation---that control and treatment groups are balanced in terms of both observed and unobserved characteristics. As a result, empirical researchers can employ straight-forward estimation strategies such as regression, inverse propensity weighting (IPW) \citep{Rosenbaum1983,Robins1994,Bang2005,Cao2009} and matching \citep{Rubin1973,Hirano2003,Abadie2016} to consistently estimate causal effects. Therefore, RCTs are widely used in clinical settings and regulators, such as the U.S.~Food and Drug Administration (FDA) and the European Medicines Agency (EMA), regard them as the gold standard to demonstrate the efficacy of a proposed treatment.

Conversely, observational data, also sometimes known as real-world data (RWD), are collected without designed interventions, such as from electronic health records or online user behavior. Despite their abundance in the Big Data era, the absence of randomization introduces bias that cannot be eliminated by increasing sample sizes, making them traditionally less favored for medical and healthcare research.


Nonetheless, RCTs are not without drawbacks, being expensive and time-consuming, which limits their scale and responsiveness. The COVID-19 pandemic highlighted the value of timely and reliable RWD, which served as crucial evidence for determining the safety and effectiveness of drugs and vaccines \citep{hansen2021assessment,pottegaard2020considerations}. 
Additionally, for rare diseases such as  male breast cancer, it may be infeasible to recruit enough patients needed to conduct clinical trials with adequate statistical power \citep{wedam2020fda}. In such cases, rigorous analyses of reliable RWD can provide valuable insights.

The concept of data fusion \citep{bareinboim2016causal}, combining observational and experimental data, presents a promising approach to harvesting the best of both realms. The combination is expected to leverage the internal validity
of experimental data and complement it with the richness of the observational data, offering a balanced approach that could potentially improve the efficiency of causal estimations through a trade-off between bias and variance.

In this paper, we present a novel power likelihood approach for effectively augmenting RCTs with observational data to improve the estimation of heterogeneous treatment effects. The remainder of this paper is organized as follows: Section \ref{sec:litreview} reviews data fusion methods proposed in recent literature. In Section~\ref{sec:setup} we outline the problem setup and main assumptions, before moving on to our proposed power likelihood approach in Section~\ref{sec:powerlikelihood}. In Section~\ref{sec:sim} we conduct a simulation study to illustrate the effectiveness  of our approach.
Section \ref{sec:pioneer6} features a comprehensive real-world data fusion study, applying our method to augmenting the PIONEER 6 semaglutide safety trial with health insurance claims data to showcase its applicability and practical benefits.  We conclude with a discussion in the final section. The code to reproduce the simulations, real data study and figures is available on 
Github: \url{https://github.com/XiLinStats/ManyData}.


\section{Related Literature}\label{sec:litreview}


In recent years, combining randomized and observational data for causal inference has attracted a lot of attention among researchers. 
Central to the data fusion problem is the existence of unmeasured confounding in the RWD, which will bias the causal effect estimation. Some methods \citep[e.g.][]{rosenman2022propensity} require unconfoundedness. Other methods make weaker assumptions about the hidden confounding:
\cite{Kallus2018} and \cite{Yang20201}  assume that the effect of unmeasured confouding on the potential outcomes follows a parametric model; \cite{athey2019surrogate,Athey2020} assume that the effect of unmeasured confounding can be fully accounted for through a surrogate outcome; \cite{YangDing2020} assume that the common causes of the treatment and outcome that are hidden in the RWD, are measured in the RCT instead.

  While the abovementioned assumptions about unmeasured confounding are reasonable, they are not testable using RWD alone. This motivates methods that do not rely on knowledge of the confounding structure. 

Formalized in \cite{viele2014use}, test-then-pool approaches are usually used to include historical controls. They start with the null hypothesis of equality between the trial and external controls and only pool the data together if this is not rejected. Typically, test-then-pool approaches require the researcher to specify the size of the test, which can be subjective and challenging in practice. \cite{Yang20201} proposed a data-adaptive approach to dynamically choose the threshold for the test static based on an estimation of the bias. A major issue with test-then-pool approaches is that when the experimental data has relatively few samples, the hypothesis test for any discrepancy is under-powered and rarely gets rejected; this results in the observational data being pooled even if it is biased. 

Other methods focus on optimizing the the tradeoff between bias and efficiency gain, quantified by mean squared error (MSE). \cite{Rosenman2020} extended the results in \cite{green2005improved} from the James-Stein shrinkage literature \citep{stein1956inadmissibility} and proposed shrinkage-styled estimators for strata-specific treatment effects. A weighted linear combination strategy has also been popular in recent literature \citep{cheng2021adaptive,Chen2021,Oberst2022}. Such estimators take a weighted average of the estimates from the RCT and RWD with the weight chosen to minimize the MSE. The key challenge is to estimate the bias of the RWD estimate and these methods broadly use the difference between the RCT and RWD estimates as an estimate of bias. \cite{dang2022cross} proposed an `experiment selector' to dynamically incorporate the RWD by stratum if its inclusion is expected to reduce MSE, estimated via cross-validation. They propose to improve the bias estimation using additional information in a negative control outcome (NCO), if one is available.

Bayesian dynamic borrowing is a set of methods are widely applied to incorporate historical studies to construct an informative prior for parameters for the outcomes or the treatment effect. The Bayesian framework offers a natural mechanism to discount external information that may be in conflict with the trial data. 
\cite{ibrahim2000power} introduced a general power prior approach that raises the likelihood of historical data to a power $\eta$, which controls the discounting of external evidence. In their proposed approach, $\eta$ is a hyperparameter and is assumed to be known, for example, set by expert opinion; obtaining a suitable value in practice can be challenging, however. As an attempt to adress this challenge, the authors, as well as \cite{duan2006evaluating} and \cite{neuenschwander2009note}, extended this formation with a hierarchical specification where the uncertainty about $\eta$ is endogenized through a prior distribution for $\eta$. A weakness of the power prior approach and its variations is that the \textit{commensurability} of the external and current data is not explicitly parameterized. Ideally, we expect the external data is considered with more weight if it appears more compatible to the current data. To this end, \cite{hobbs2011hierarchical,hobbs2012commensurate} introduced a location commensurate power prior (LCPP), where the different parameters in the current and historical data are explicitly modeled to be normally distributed, with additional priors chosen for the paramters of the normal distribution. Extending the commensurate prior approach, \cite{schmidli2014robust} proposed a meta-analytic-predictive posterior approach to borrow information from multiple historical trials where a mixture of standard priors is used.

Despite the advances in Bayesian dynamic borrowing, the challenge of specifying hyperparameters or commensurability functions persists, and applications primarily focus on aggregate data or with overly simplistic data-generating models. Our proposed power likelihood method enhances the existing Bayesian dynamic borrowing literature in two significant ways.  First, it eliminates the need for analysts and experts to specify a learning parameter or commensurability function, by adopting a data-adptive approach to determine the optimal learning rate.  Second, it facilitates application to individual-level participant data, with covariates incorporated through complex and realistic data-generating models via the frugal parameterization \citep{Evans2021}.

Compared to existing methods that are either limited to incorporating only external controls \citep{schuler2020increasing} or require the presence of both treatment and control arms in the real-world data \citep{Kallus2018,Oberst2022,cheng2021adaptive,Rosenman2020,Yang20201}, our proposed approach offers the added flexibility of integrating external data either from the control arm alone or from both treatment and control arms. This flexibility is valuable because the treatments under study, such as a new drug or a social policy, may not yet be approved or implemented in the real world. Being able to leverage the real-world data, even without treatment arm observations, is important. Additionally, if a comparable treatment is already prevalent in the real world, utilizing information from both arms will make the data fusion more effective.

\section{Setup, Notation and Assumptions}\label{sec:setup}


\subsection{Setup and notations}
We consider two data sources, experimental data $\mathcal{D}_e$ of size $n_e$ and observational data $\mathcal{D}_o$ of size $n_o$. $S \in \{0,1\}$ denotes whether an observation is from the RCT ($S = 1$) or the observational data ($S = 0)$.
The causal models for the two data sources are represented by the graphs in Figure \ref{fig:DAG}, where $T \in\{0,1\} $ represents a binary treatment, $\bs W \in \mathbb{R}^d$  is a vector of $d$ pre-treatment covariates, $Y\in \mathbb{R}$ is the outcome of interest, and $U$ represents unmeasured confounding.  In both data sources, we assume that the observations $X_i = \left(Y_i,T_i,\bs W_i \right)$ are independently and identically distributed, respectively. We denote random variables with capital letters, like $\bs X$ and $Y$, and their instantiations with corresponding lowercase letters, like $\bs x$ and $y$. 

Following the potential outcomes framework \citep{rubin1974estimating,neyman1923applications} to represent causal relationships, we maintain the standard Stable Unit 
of Treatment Value Assumption (SUTVA) and let $Y_i(t)$ be the potential outcomes for unit $i$ had it factually received treatment $t$, and the observed outcome $Y_i = Y_i(t)$ if $T_i = t$. The treatment effect for each individual (ITE), $\tau_i$, is defined as a contrast of potential outcomes. For a continuous outcome, it is customary to define the ITE as the difference between potential outcomes $Y(1) - Y(0)$. For a binary outcome, candidate definitions of the treatment effect include risk difference (RD), $Y(1) - Y(0) $, relative risk $\frac{Y(1)}{Y(0)}$ and odds ratio (OR) $\frac{Y(1)/1-Y(1)}{Y(0)/1-Y(0)}$. Our proposed method is applicable to all of these definitions and---for illustration purposes---we adopt the difference in potential outcomes to define causal effect. If we average the ITE over a target population, we get the average treatment effect (ATE), $\tau \equiv \E[Y(1) - Y(0)] $. 

In practice, there are scenarios where we are interested in how the treatment effect varies in subgroups. Based on the setup depicted in Figure~\ref{fig:DAG}, we can further divide the set of observed covariates $\bs W$ into two sets, $\bs Z$ and $\bs C$. The set $\bs C \subseteq \bs W$ denotes the causal effect modifiers we are interested in, and $\bs Z  = \bs W \setminus \bs C$ represents the remaining covariates that we want to marginalize over. 
Our target estimand is the conditional treatment effect (CATE) $\tau(\bs c) \equiv \E[Y(1) - Y(0) \mid \bs C = \bs c]$, which refers to the average treatment effect given $\bs {C}=\bs {c}$, but marginal over values of $\bs Z$ in the target population. It is worth highlighting that the choice of variables into set $\bs C$ depends on the specific causal question of interest; the ATE can be seen as a special case where the conditioning set $\bs C$ is empty.



\begin{exm}\label{exm}
We use an example to further illustrate our setup with the distinction of covariate sets $\bs C$ and $\bs Z$. In a medical context, clinicians may be interested in the efficacy of a drug on patients with certain characteristics. Suppose we have access to an RCT and an observational dataset, each of which record whether individuals took the drug $T$, some health outcome $Y$, body mass index (BMI) $C$ and blood glucose level $Z$. 
We assume that both $C$ and $Z$ influence the outcome $Y$. In addition, there is a link between BMI $C$ and blood glucose level $Z$.
To give guidance on the prescription of this drug, we are interested in the causal effect moderated by BMI $C$ and want to marginalize it over the distribution of $Z$. Specifically, we are interested in $ \tau ( {c}) = \E \left [ \E \left[Y(1) - Y(0) \cmid C =  c, Z\right]\right]$.
We will revisit this example in Section~\ref{sec:sim}, where we conduct a simulation study following this setup.
\end{exm}




\begin{figure}{\label{1:DAG}}
 \begin{center}
 \begin{tikzpicture}
[node distance=20mm, >=stealth, scale=3]
 \pgfsetarrows{latex-latex};
 \begin{scope}
 \node[rv]  (1)              {$W$};
 \node[rv, right of=1] (2) {$T$};
 \node[rv, right of=2] (3) {$Y$};
 \node[rv, red,above of=2] (4) {$U$};

 \draw[deg] (1) -- (2);
 \draw[deg] (2) -- (3);
 \draw[deg] (1) to[bend right] (3);
 \draw[degl] (4) to (3);
 \node[below of=2, xshift=0mm, yshift=0mm] {(a)};
 \end{scope}
 \begin{scope}[xshift=2cm]
 \node[rv]  (1)              {$W$};
 \node[rv, right of=1] (2) {$T$};
 \node[rv, right of=2] (3) {$Y$};
 \node[rv, above of=2] (4) {$U$};

 \draw[deg] (2) -- (3);
 \draw[degl] (4) -- (2);
 \draw[deg] (1) -- (2);
 \draw[deg] (1) to[bend right] (3);
 \draw[degl] (4) to (3);

 \node[below of=2, xshift=0mm, yshift=0mm] {(b)};
 \end{scope}
 \end{tikzpicture}
 \caption{Causal models for (a) the experimental data $\mathcal{D}_e$  and (b) the observational data $\mathcal{D}_o$. 
 }
 \label{fig:DAG}
 \end{center}
\end{figure}

\subsection{Assumptions}
We now state our main assumptions.

\begin{assumption}\label{assump:rct_ignorability} (RCT ignorability)
 (i) \textbf{Unconfoundedness} $Y(t) \indep T \cmid W, S = 1$ for $t \in \{0,1\}$  (ii) \textbf{Positivity} $0 < \mathbb P(T =1 \cmid W, S = 1) < 1 $ for all ($W, S=1$).
\end{assumption}

 Assumption \ref{assump:rct_ignorability} (i) states that there is no unmeasured confounding and treatment assignment is independent of potential outcomes. This assumption is satified when treatment is randomly assigned. Assumption \ref{assump:rct_ignorability} (ii) ensures that for each individual, there is a nonzero probability of being assigned to either treatment or control group. This ignorability assumption is inherently satisfied in a well-conducted RCT. 

\begin{assumption}\label{assump:rct_representativeness} (RCT representativeness)
$\E[Y(1) - Y(0) \cmid \bs C, S = 1]$ = $\tau(\bs c)$.
\end{assumption}

Assumption \ref{assump:rct_representativeness} maintains that the RCT sample is representative, allowing the CATE of interest to be transportable from the RCT to the target population. This assumption allows us to to focus on gaining efficiency as the primary goal of data fusion, without concerning with transporting or generalizing the inference to a different population. 


\begin{assumption}\label{assump:transport}(RWD transportability)
 $\E[Y(1) - Y(0) \cmid C, S = 0] =  \E[Y(1) - Y(0) \cmid C, S = 1]$.
 \end{assumption}

Assumption \ref{assump:transport} states that the CATE  of interest in the RWD is identical to the CATE in the RCT population. It is important to highlight that Assumption \ref{assump:transport} does not stipulate the mean of the potential outcomes to be the same across $\mathcal{D}_e$ and $\mathcal{D}_o$. Therefore, Assumption \ref{assump:transport} is strictly weaker than assuming  $\E[Y(t) \cmid S = 0] =  \E[Y(t) \cmid S = 1]$ for $t \in (0,1)$; this gives flexibility when the absolute level of the outcome differs between the RCT and RWD, for reasons such as different time windows, regions or standards of care. In practice, to improve adherence to this assumption, we can follow the  target trial emulation framework, as outlined in \cite{hernan2016using} and \cite{hernan2022target}. This approach systematically emulates the design of a target RCT using observational data, thereby enhancing the comparability between the RCT and RWD, as demonstrated in the PIONEER 6 data study in Section \ref{sec:pioneer6}.

\section{A Power Likelihood Approach}\label{sec:powerlikelihood}


Traditional Bayesian inference assumes that the data model, $f(x;\phi)$ is correct up to the unknown parameter value $\phi$. However, Bayesian inference loses its predictive optimality \citep{zellner1988optimal} when the data models are misspecified. \cite{bissiri2016general} proposed a framework for \textit{general Bayesian inference}, where parameters are connected to observations through a loss function other than the traditional likelihood function:  
$$
\pi(\phi\mid x)\propto \exp \left \{-l(\phi,x) \right\}\pi(\phi),
$$
where $l(\phi,x)$ is a loss function; this relaxes the requirement of a `true' data-generating mechanism. This is relevant to our data fusion problem because by omitting the unmeasured confounder $U$, the model for the observational data is inevitably misspecified. 
Under this overarching framework, a popular solution to robustly allow for Bayesian learning under model misspecification is to raise the likelihood to a fractional power. That is, $\pi_\eta(\phi\mid X ) \propto \pi(\phi)\prod_{i=1}^n f(X_i;\phi)^\eta $, indexed by a power $\eta$. This approach is described as a ``power prior" in \cite{ibrahim2000power}, a ``data-modified prior"  in \cite{walker2001bayesian}, and a ``power likelihood" in \cite{holmes2017assigning}.

Adapting to our inference problem, we take the joint likelihood but only raise the likelihood of the observational data to a power $\eta$:
$$
f_\eta(\bs{X_e},\bs{X_o}; \phi) = f_e(\bs{X_e}; \phi) \times f_o \left(\bs{X_o};\phi \right)^\eta,
$$
where $\bs{X_e}$ and $\bs{X_o}$ represent the observations from $\mathcal{D}_e$ and $\mathcal{D}_o$, and  $ f_e(\bs{X_e};\phi)$ and $f_o\left(\bs{X_o};\phi \right)$ are the joint densities respectively. This is equivalent to, under the \textit{general Bayesian framework}, defining a loss function $ l_\eta(\phi; \bs{X_e},\bs{X_o} ) = - \left\{\log f_e\left (\bs{X_e};\phi \right ) + \eta \log f_o\left(\bs{X_o};\phi \right) \right\}$. Then by Bayes' rule, the posterior distribution of $\phi$ becomes:
\begin{align*}
    \pi\left(\phi \mid \bs{X_e},\bs{X_o}\right) &\propto \exp\left \{-l_\eta \left(\phi;\bs{X_e},\bs{X_o} \right)\right \} \pi(\phi)\\
    & = f_e(\bs{X_e};\phi) \cdot f_o\left(\bs{X_o};\phi\right)^\eta  \cdot \pi(\phi),\stepcounter{equation}\tag{\theequation}\label{eq:Bayes}
\end{align*}
where $\pi(\phi)$ is the prior distribution. We can see that the parameter $\eta$ works like a dial moderating the influence from the observational data, relative to that from the experimental data. When $\eta = 0$, the loss function is just the negative log likelihood of the experimental data $\mathcal{D}_e$ and the influence of the observational data is cut off as if $\mathcal{D}_o$ is excluded from the analysis. When $\eta = 1$ then it is the conventional Bayesian posterior and means that we treat $\mathcal{D}_e$ and $\mathcal{D}_o$ equally for inference. Intuitively, we want to choose $\eta$ to be between $0$ and $1$.

Two important components of the power likelihood approach are the likelihood and the power $\eta$. In the remainder of this section, we discuss the likelihood we propose to use and the robust selection of $\eta$.

\subsection{Frugal parameterization}\label{sec:frugal}

To specify the joint densities $ p_e$ and $p_o$ in \eqref{eq:Bayes}, we propose to adopt the \textit{frugal parameterization} \citep{Evans2021}, which is designed specifically for causal inference applications. Following its framework, we can break down a joint distribution $p(\bs {z},t,y \cmid \bs {c})$ into three separate pieces:
\begin{enumerate}[label = (\alph*)]
    \item the distribution of treatment and pre-treatment covariates: $p_{\bs Z T | \bs C }( \bs z, t \cmid \bs c)$
    \item  the causal quantity of interest: $p_{Y\mid T \bs C}^*(y \mid t, \bs c):=P(Y(t)=y \mid  \bs C = \bs c )$, and
    \item  a dependence measure $\varphi^*_{Y \bs{Z} | T \bs C} $  between potential outcomes $Y(t)$ and $\bs {Z}$ conditional on $\bs C$.
\end{enumerate}
Examples of objects indicated in (c) include copulas and  conditional odds ratios. Consistent with notations in \citep{Evans2021}, we use an asterisk to denote causal or interventional distributions where the treatment is set to value $t$ through intervention.
The advantages of using frugal parameterization are three-fold. Firstly, it gives a likelihood using a causally-relevant parameterization to the observational data. Secondly, it isolates the causal quantity of interest, $p_{Y(t)\mid \bs C}^*$, from the rest of the joint distribution, which means that we can directly target inference for this quantity, and treat the others parameters as a nuisance model. Finally, the frugal parameterization allows the non-causal distributions to differ between the RCT and observational data, for example, the joint distribution of covariates and treatment assignment, 
$p_{\bs Z T | \bs C }( \bs z, t \cmid \bs c)$.\par 
Under the \textit{frugal parameterization}, we can factorize joint densities $p_e$ and $p_o$ as below:
\begin{align}
    p_e(y ,\bs{z}, t \cmid \bs{c};\theta,\gamma ) &= p_{e,\bs ZT | \bs{C}}(\bs z,t \cmid \bs{c};\gamma) \cdot p_{Y\mid T \bs {C}}^*(y \mid t, \bs {c} ; \theta) \cdot \varphi^*_{e,Y\bs {Z}\mid T\bs{C}}(y, \bs {z} \cmid t, \bs {c}; \gamma ) \stepcounter{equation}\tag{\theequation}\label{eq:frugal1}\\
    p_o(y ,\bs{z}, t \cmid \bs{c} ;\theta,\psi ) &= p_{o,\bs ZT | \bs{C}}(\bs z,t \cmid \bs{c};\psi) \cdot p_{Y\mid T \bs {C}}^*(y \mid t, \bs {c} ; \theta) \cdot \varphi^*_{o,Y\bs {Z} \mid T\bs{C}}(y,\bs {z} \cmid t, \bs {c} ; \psi )\stepcounter{equation}\tag{\theequation}\label{eq:frugal2},
\end{align}
where we decompose the set of parameters $\phi$ into three subsets $\phi = (\gamma, \theta,\psi)$; here $\theta$ is the set of shared causal parameters, while $\gamma$ and $\psi$ parameterize the nuisance models for the RCT and RWD respectively. Additionally, to ensure a consistent estimation of the HTE function $\tau(\bs c)$, we further require a correctly specified model for $p_{e,Y\mid T\bs {C}}^*(y \mid t, \bs {c} ; \theta)$.

\begin{assumption}\label{assump:correctspec}
The distribution of potential outcomes  $P(Y(t) \mid \bs C)$ in the randomized data  $\mathcal{D}_e$ is correctly specified by a parametric model $p_{e,Y\mid T\bs {C}}^*(y \mid t, \bs {c} ; \theta)$.
\end{assumption}

A convenient choice of such specification is a linear model, although linearity is not required by our proposed method. Unlike requiring a correctly specified data model for the full joint distribution, Assumption~\ref{assump:correctspec} is strictly less stringent. Furthermore, it imposes a requirement only on the randomized data, but makes no restriction on the observational data model. 

\begin{rmk}
With Assumption~\ref{assump:correctspec}, our inference takes place in the so-called $\mathcal{M}$-closed world. However, as famously stated by \cite{box1976science}, ``all models
are wrong.'', the assumption of a correct specification is unlikely to hold in practice. Instead, we conduct causal inference in the $\mathcal{M}$-open world \citep{bernardo1994bayesian}, where maximum likelihood estimation finds the element of the assumed model which minimizes the Kullback-Leibler (KL) divergence to the data. Additionally, it is worth noting that, in the case of misspecification, if treatment is binary, the ATE estimate constructed through marginalizing over all conditioning variables $\bs C$ will nonetheless be consistent.
\end{rmk}


We now apply this parameterization to specify the power posterior in (\ref{eq:Bayes}). Substituting the joint densities $p_e(\bs{X_e};\theta,\gamma)$ and $p_o\left(\bs{X_o};\theta,\psi\right)$ by  (\ref{eq:frugal1}) and (\ref{eq:frugal2}) gives
\begin{align*}
     \pi_\eta(\theta, \psi, \gamma \mid \bs{X_e},\bs{X_o}) &\propto  p_e(\bs{X_e};\theta,\gamma) \cdot p_o(\bs{X_o};\theta,\psi
     )^\eta \cdot  \pi(\theta, \psi,\gamma)\\
     & \propto  p_{e,\bs ZT \cmid \bs{C}}\left(\bs z,t \cmid \bs{c};\gamma\right) \cdot p_{Y\mid T\bs {C}}^*(y \mid t, \bs {c} ; \theta) \cdot \theta^*_{e,Y\bs {Z}\mid T\bs{C}}\left (y, \bs {z} \cmid t, \bs {c}; \gamma \right) \\
     &\quad\times \left(p_{o,\bs ZT \cmid \bs{C}}\left(\bs z,t \cmid \bs{c};\psi\right) \cdot p_{Y\mid T\bs {C}}^*(y \mid t, \bs {c} ; \theta) \cdot \theta^*_{o,Y\bs {Z} \mid T\bs{C}}\left(y,\bs {z} \cmid t, \bs {c} ; \psi \right) \right)^\eta  \\
     &\quad \times \pi(\theta, \psi,\gamma). \stepcounter{equation}\tag{\theequation}\label{eq:Bayes2}
\end{align*}

\subsection{Choosing the optimal influence factor $\eta$}
We can see that $\eta$ plays a crucial role in controlling how much information we want to borrow from the observational data. The selection of $\eta$ needs to be performed carefully: the observational data can be useful in reducing the variance, however, we do not want to contaminate our causal estimate with excessive bias by setting $\eta$ too high. Ideally, we want $\eta$ to be chosen in a robust manner that is adaptive to the data, where we incorporate more information when the intrinsic bias in the  estimate from the observational data, $\hat\theta_o$, is small and its variability is low, and vice versa. Specifically, we use MSE as the loss function, and we aim at minimizing the risk of our resultant estimator 
$R(\hat{\theta}_\eta,\theta) = \E \|\hat{\theta}_\eta-\theta \|^2 $.

In recent literature, several methods for selecting the $\eta$ in the power likelihood have been proposed, mainly in the context of addressing model misspecification; for example, the \textit{SafeBayes} algorithm \citep{grunwald2017inconsistency}, expected information matching \citep{holmes2017assigning,lyddon2019general}, and frequentist coverage probability calibration \citep{syring2019calibrating}. A review and empirical comparison of these methods can be found in \cite{wu2020comparison}.

However, these methods primarily address misspecifications where the posterior is consistent and concentrated around the true values \citep{holmes2017assigning,syring2019calibrating,lyddon2019general}, or deal with 'benign' misspecifications related to heteroskedasticity \citep{grunwald2017inconsistency}. These scenarios do not apply to our data fusion setting, where the bias caused by unobserved confounding does not diminish as the sample size increases. 

Considering our problem setting and objectives, similar to the choice of \cite{carmona2020semi}, we propose to select $\eta$ by maximizing the expected log point-wise predictive density (ELPD): 
\begin{align*}
    \operatorname{ELPD}(\eta) &= \E_{\Tilde{\X}} \log p_\eta(\Tilde X \cmid \bs x_e , \bs x_o)
    =  \int_{\Tilde{\X}} p_t(\Tilde x) \log p_\eta(\Tilde x \cmid \bs x_e , \bs x_o) \, d \tilde x\stepcounter{equation}\tag{\theequation}\label{eq:elpd},
\end{align*}
where
$$
p_\eta(\Tilde x \cmid \bs x_e , \bs x_o) = \int_{\Gamma}\int_\theta p(\Tilde x \cmid \theta, \gamma) p(\theta, \gamma \cmid  \bs x_e , \bs x_o) \,  d\theta \, d\gamma  
$$
is the posterior predictive distribution indexed by $\eta$, and the expectation is with respect to the `true' data generating process $p_t$. 
As this `true' distribution is unknown, there are different ways to estimate ELPD. A commonly used method to approximate it is leave-one-out cross-validation (LOO), which involves leaving one observation out at a time, evaluating the posterior predictive density on this observation, and then averaging over all observations. This method is computationally expensive as one can imagine. \cite{vehtari2017practical} introduced an efficient computation of LOO using Pareto-smoothed importance sampling (LOO-PSIS) which avoids repeated partition of the dataset.

Another estimation method is to use the widely applicable information criterion (WAIC) which is proven to be asymptotically equal to LOO \citep{watanabe2010asymptotic}. The WAIC method simply uses the ordinary posterior to estimate the density of each observation, and subtracts the effective number of parameters, defined as
\begin{align*}
\widehat{\operatorname{ELPD}}(\eta) &= \frac{1}{n_e}\sum_{i=1}^{n_e} \log \hat{p}_\eta(x_i \cmid \bs x) - \hat{d}_{\text{WAIC}},
\end{align*}
where $\hat{p}_\eta(x_i \cmid \bs x)$ is estimated using the posterior samples of the parameters, and $\hat{d}_{\text{WAIC}}$ is the estimated effective number of parameters. \cite{gelman1995bayesian} provide a mean-based and a variance-based definition for $\hat{d}_{\text{WAIC}}$, and consistent with \cite{vehtari2017practical}, we use the variance-based definition $d_{\text {WAIC }}=\sum_{i=1}^{n_e} \var_{\text {post}}\left\{\log p(x_i \mid \theta, \gamma)\right\}$, where $\var_{\text{post}}$ is the posterior variance of the log predictive density for each data point $x_i$ in the randomized data.
We propose to evaluate the ELPD on $\mathcal{D}_e$, as the randomization ensures the causal relationship in the experimental data is correct. We then apply a grid search in $[0,1]$ to find the $\eta^*$ that maximizes
$\widehat{\operatorname{ELPD}}(\eta^*)$.
\begin{rmk}\label{rmk:elpd}
It is worth noting the connection between ELPD and the KL divergence measure. We can rewrite (\ref{eq:elpd}) as 
\begin{align*}
    \operatorname{ELPD}(\eta) &=  \int_{\X} p_t(\Tilde x) \log p_\eta(\Tilde x \cmid \bs x_e , \bs x_o)  \, d \tilde x\\
    &= \int_{\X}  p_t(\Tilde x) \left\{\log \frac{ p_\eta(\Tilde x \cmid \bs x_e , \bs x_o)}{p_t(\Tilde x)} + \log p_t(\Tilde x)\right\} \, d \tilde x\\
    & = \int_{\X}  p_t(\Tilde x)\log \frac{ p_\eta(\Tilde x \cmid \bs x_e , \bs x_o)}{p_t(\Tilde x)} \, d \tilde x + \int_{\X}  p_t(\Tilde x) \log p_t(\Tilde x) \, d \tilde x\\
    &= - d_{\operatorname{KL}} \left(p_t \cmid p_\eta \right) - H(p_t)\stepcounter{equation}\tag{\theequation}\label{eq:elpd2},
\end{align*}
where $H(p_t) = -\int_{\X}  p_t(\Tilde x) \log p_t(\Tilde x) \, d \tilde x$ is the entropy of $p_t$, a constant independent of $\eta$. Essentially, choosing the $\eta$ via maximizing ELPD is equivalent to selecting the posterior predictive distribution, indexed by $\eta$, that is closest to the true data distribution in KL-divergence. This confirms that ELPD is a proper scoring rule.
\end{rmk}

Our proposed method to select $\eta$ is summarized in Algorithm 1.

\begin{algorithm}
	\caption{Power likelihood $\eta$ selection} 
	\begin{algorithmic}
	    \State Initialize $\eta^* \leftarrow 0$ ,  $\operatorname{ELPD}^* \leftarrow -\infty$ 
		\For {$i=0,1,2,\ldots,N$}
		    \State Let $\eta = i/N$
			\State Sample $(\theta^{(i)},\psi^{(i)},\gamma^{(i)}) \sim \pi_\eta\left(\theta, \psi, \gamma \mid \bs X_e,\bs X_o\right)$ using any appropriate sampler
			\State Compute $\widehat{\operatorname{ELPD}}(\eta)$ using posterior samples $(\theta^{(i)},\psi^{(i)},\gamma^{(i)})$
			\If{$\widehat{\operatorname{ELPD}}(\eta) > \operatorname{ELPD} ^*$}
			\State $\operatorname{ELPD} ^* \leftarrow \widehat{\operatorname{ELPD}}(\eta) $
			\State $\eta^* \leftarrow \eta$ 
			\EndIf
		\EndFor
	\State \Return {$\eta^*$}
	\end{algorithmic} 
\end{algorithm}
Once an optimal $\eta$ is selected, we can then perform likelihood-based inference.

\subsection{Normal approximations to the posterior distribution}\label{sec:normalapprox}

The Bernstein-Von Mises theorem states that in a smooth, finite-dimensional model, as the sample size $n$ approaches infinity, the posterior distribution of a parameter $\phi$ will converge to a normal distribution with mean $\hat \phi_n$ being the maximum likelihood estimator, and variance $\mathcal{I}_{\phi_0}^{-1}$, where $\mathcal{I}_{\phi_0}$ is the Fisher information matrix of the model at the true value $\phi_0$ \citep{van2000asymptotic}. When the model is correctly specified, $\hat \phi_n$ converges to the true value $\phi_0$ at the rate of $\sqrt{n}$. For a misspecfied model, $\hat \phi_n$ instead converges to $\tilde \phi$, which minimizes the KL divergence from the model to the truth.

In our context, for the combined parameter vector $\phi = (\theta, \gamma, \psi)^T \in \mathbb{R} ^{m \times 1}$,  where $m = p+d+k$ is the total dimension, the theorem leads to the following lemma:

\begin{lem}\label{lem:norm_approx}


Given the the MLE of $\phi$ in $\mathcal{D}_e$ and $\mathcal{D}_o$ as $\hat \phi_e$ and $\hat \phi_o$ respectively, both are asymptotically normal with $\sqrt{n_e}(\hat\phi_e - \phi^*) \to N(0, V)$ and $\sqrt{n_o}(\hat\phi_o - \phi^* - \delta) \to N(0, W)$, where $\phi^*$ is the true value of $\phi$, $\delta$ represents the `bias' and $V,W \in \mathbb{R}^{m \times m}$.
For $\hat{\phi}_\eta$, the MLE under the $\eta$-powered likelihood,
we find
$
\hat \phi_\eta =  \left(n_eV^{-1} + \eta n_o W^{-1}\right)^{-1}\left(n_eV^{-1}\hat\phi_e + \eta n_o W^{-1}\hat\phi_o \right)
$ and that it has asymptotic variance $V^{-1} + \eta W^{-1}$.
\end{lem}
Derivation of these quantities can be found in Appendix \ref{sec:app_norm}.  A consequence of the lemma is that we can approximate the posterior of $\phi$ using a normal distribution with mean $\hat\phi_\eta$ and variance $\mathcal{I}(\hat \phi_\eta)^{-1} := V^{-1} + \eta W^{-1}$.  In simulations presented in Section~\ref{sec:sim}, we use the sandwich estimators of $V$ and $W$ and Fisher information calculated using the R package \texttt{causl} \citep{causl}. 

Our proposed method, as described in Algorithm 1, involves iterative sampling from posterior distributions to find the optimal $\eta$. While Markov chain Monte Carlo (MCMC) is a standard approach for posterior sampling, it becomes computationally challenging for large datasets, such as we expect for $\mathcal{D}_o$. This is exacerbated when the parameter vector $\phi$ is high dimensional, which challenges sampling efficiency. Lemma \ref{lem:norm_approx} provides a solution for such cases. A nice property of this asymptotic approximation is that the bigger the data sets, the better the approximation, which will clear the computational hurdle and make our method scalable for application to big data.

\begin{rmk}
    In instances where there are no observational subjects receiving treatment, the corresponding entries in the Fisher information matrix  $\mathcal{I}(\hat \phi_o)$ are zero, indicating that it the provides no information on the coefficients for treatment-related terms in the $p^*_{Y | T C} (y \cmid t, c)$ model  component. 
\end{rmk}

\subsection{Asymptotics}
Intuitively, the $\hat \eta$ that maximizes ELPD in (\ref{eq:elpd2}) depends on the magnitude of bias introduced by the observational dataset. The more biased $\D_o$ is, the less we should incorporate its influence. We are particularly interested in studying the asymptotics of $\hat \eta$ and the resulting estimator $\hat \theta_{\hat \eta}$. Similar to the discussion in \cite{YangDing2020} and \cite{dang2022cross}, we assume that the bias in the observational data $\D_o$ due to hidden confounding is not fixed but depends on its sample size, even though this is not plausible in practice.  Specifically, we consider two regimes of local alternative based on the rate of convergence of the bias to zero.


\begin{thm}[\textbf{Consistency}]\label{thm:consistency}
Let $\delta = \delta(n_o) = \delta^*/n_o^k$ denote the bias in $\D_o$.
%
\begin{enumerate}[label = (\alph*)]
    \item If $0 \leq k < 1/2$ and $\delta^* \neq 0$, then $\hat \eta$ converges to 0, and $\hat \phi_{\hat \eta} = \phi^*+ o_p(n^{-1/2})$. 
    \item If $k \geq 1/2$ or $\delta^* = 0$ then $\hat \eta $ does not converge.

\end{enumerate}
\end{thm}

Theorem 1 suggests that when bias is asymptotically (statistically) non-negligible, the selected influence factor $\hat \eta$ converges to 0 and the estimator $\phi_{\hat \eta}$ is $\sqrt{n}$-consistent. As a result, the MSE of the estimator $\phi_{\hat \eta}$ converges to that from using the experimental data only. To provide intuition for this theorem, as $n_o$ grows, the fraction of its influence we want to incorporate reduces and as $n_e$ grows, the need to augment it with external data diminishes at the same time. However, when the observational data has asymptotically negligible bias, the selected $\hat \eta$ will not converge to zero. On the contrary, as bias diminishes close to zero, we will very likely infer that $\hat \eta = 1$ is the best value as we restrict $\eta$ to be between 0 and 1. The proof is presented in Appendix \ref{sec:app_proof}.

\section{Simulation}\label{sec:sim}

            
    

This section presents simulations that empirically demonstrate how our method robustly moderates the influence of the observational data through the $\eta$-powered likelihood, thereby providing adaptive inference results across various degrees of unmeasured confounding in finite samples.
\subsection{Simulation setup}
We followed the setup outlined in Example~\ref{exm}, observing two pre-treatment covariates: an obesity indicator  $C$ and a continuous measure of creatinine level $Z$; a binary treatment variable $T$ indicating drug administration; and a continuous outcome $Y$ which measures change in average blood glucouse level. Additionally there exists an unobserved confounder $U$, affecting both the treatment and the outcome. The influence of $U$ means that any causal estimate based on the assumption of unconfoundedness in the observational data will be biased. Specifically, we simulate the observational data $\mathcal{D}_o$ from the following data models:
\begin{align*}
    U&\sim \Bernoulli(0.5) &   C&\sim \Bernoulli(0.5)  & Z \mid C &\sim N(\mu_z, \, 1) \\
    T \mid C,Z,U &\sim \Bernoulli(\mu_t) & Y(t) \mid C &\sim N(\mu_y, \, 1),
\end{align*}
where,
\begin{align*}
\mu_z &= 1 + \, C & \logit \mu_t & = -3  + \,  C +  \, Z + \,  C\, Z + \omega U & \mu_y &= 1 +  C + 0.1 \, T + 0.1 \, C \, T + \omega U.
\end{align*}
 Additionally, let $\varphi^*_{YZ \mid TC}$, the dependence structure between $Y$ and $Z$ given $T$ and $C$, be a conditionally bivariate Gaussian copula, with correlation parameter $\rho_{t} = 2\expit(1 + 2.5t)-1$.

We use the same parameterization for the experimental data $\mathcal{D}_e$ with two adjustments: (i) treatment is randomly assigned and hence replace the mean of $T$ with $\mu_t = 0.5$, and (ii) the outcome $Y$ is assumed independent of any hidden confounding, so we set $\omega = 0$. This simulation framework is designed to reflect a realistic data fusion scenario where, the RCT is free of confounding while there is an unmeasured factor $U$, such as the access to and quality of healthcare, that affects both treatment and outcome, as well as introducing extra variability in the outcome.

We set the sample sizes to 1,000 and 500 for the observational and experimental data, respectively. The results we show in this section are averaged across 500 sets of synthetic datasets.

\begin{rmk}
In this as well as other simulations that we have tested, we compared the ELPD estimated by LOO-PSIS and WAIC method, implemented through the \texttt{loo} package in R \citep{vehtari2017practical}, and they almost always give very close results. We use the LOO-PSIS approximation in this simulation. We have also compared obtaining posterior parameter samples using the Metropolis-Hasting within Gibbs MCMC sampler, with sampling directly from the approximated normal distribution as described in Section \ref{sec:normalapprox}. With sample sizes of $n_e=500$ and $n_o=1,\!000$ in a 12-dimensional parameter space, the two approaches lead to very similar $\eta$'s being selected. For computational efficiency, we sample directly from the approximated normal distribution in this simulation study.
\end{rmk}

\begin{figure}
    \centering
    \includegraphics[width=10cm, height = 8cm]{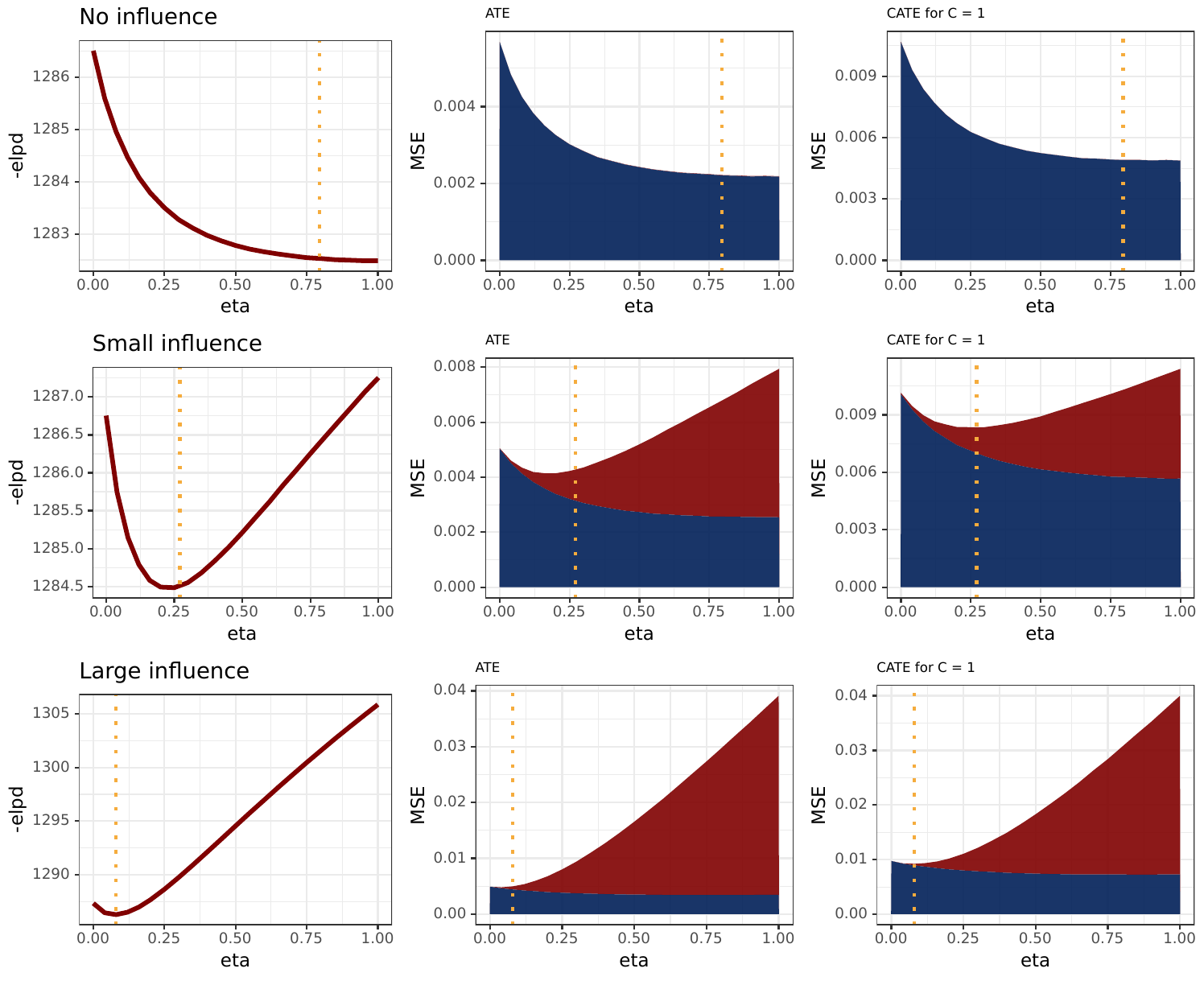}
    \caption{Simulation results for scenarios where $U$ has no influence ($\omega = 0$), small influence ($\omega = 0.5$) and large influence ($\omega = 1$) on $T$ and $Y$. Left column: average negative ELPD with $\eta \in [0,1]$. The yellow dotted lines mark the average $\eta$ selected. Middle column: MSE of the ATE estimate. Right Column: MSE of the CATE estimate in the subgroup with $C = 1$. The MSE is decomposed into bias (red-shaded area) and variance (blue-shaded area).}
    \label{fig:sim_res1}
\end{figure}

\subsection{Moderating the bias-variance trade-off}\label{sec:sim_main}
Figure~\ref{fig:sim_res1} presents three scenarios where unmeasured confounding has an influence of different magnitudes: no influence ($\omega = 0$), small influence ($\omega = 0.5$) and large influence ($\omega = 1)$. The first column shows the average negative ELPD at each value of $\eta$. Our proposed method searches for the $\eta$ that maximizes ELPD, which corresponds to the lowest point on the curve. The second and third columns show the MSE of the ATE and CATE\footnote{While we only show the CATE in the subgroup where $C = 1$, we have analyzed the MSE for CATE estimates in the subgroup $C = 0$ and it exhibited a very similar pattern of MSE loss as $\eta$ varied.} estimates evaluated on $\mathcal{D}_e$ at different values of $\eta$. We decompose MSE into variance (in blue) and squared bias (in red) to give an intuitive illustration of the moderating effect of $\eta$ on the bias-variance trade-off: as we increase $\eta$, we include more information from $\mathcal{D}_o$ and hence the variance reduces, yet at the expense of an increase in bias.

When $U$ has no influence on $T$ and $Y$, that is, $\mathcal{D}_o$ is not biased and is just `as good as' $\mathcal{D}_e$, the average $\eta$ selected is around $0.80$; this is reasonable, because in this context we want to include as much information in $\mathcal{D}_o$ as possible. Unsurprisingly, the corresponding MSEs of ATE and CATE estimates are both considerably lower than if we do not include $\mathcal{D}_o$ at all. 

When $U$ has a small influence on $T$ and $Y$, our method chooses a smaller $\eta$, averaging around $0.27$. Again, the corresponding MSEs of ATE and CATE estimates are both lower than at $\eta = 0$. When $U$ has a larger influence, that is, $\mathcal{D}_o$ is more biased, our method responds by selecting an even smaller $\eta$. Although the corresponding MSE of the ATE estimate is almost unchanged, we achieve a reduction in MSE of for the CATE estimates when compared with $\eta = 0$.  

\subsection{Inference}
By the Bernstein-von-Mises theorem, under the condition that the prior is continuous and positive in a neighborhood around the MLE, the Bayesian credible intervals approximately align with frequentist confidence intervals as the sample size approaches infinity. Nevertheless, it is important to recognize that the MLE derived from an $\eta$-powered likelihood of a combined dataset may not coincide with the MLE from the experimental data alone. Consequently, there is no theoretical assurance that the combined MLE's distribution will converge around the true parameters of the experimental data, and the credible interval may not offer the nominal coverage rates, particularly in the presence of hidden confounding. 

Despite the lack of theoretical guarantee, the empirical coverage rates presented in Table \ref{table:5_2} closely matched the nominal rate of $95\%$, for the ATE $\tau$, CATE $\tau(1)$, and $\tau(0)$, under varying levels of confounding, $\omega$. Moreover, while using the experimental data alone yields test powers for $\tau$, $\tau(1)$ and $\tau(0)$ of approximately $57\%$, $52\%$, and $18\%$ respectively, data fusion substantially enhances these figures, delivering the efficiency gain as desired. Overall, we achieve reductions in MSEs across different confounding scenarios.

\begin{table}[ht]
\centering
\begin{tabular}{ccccccccccc}
\hline
$\omega$ & $\eta$ & \multicolumn{2}{c}{ATE $\tau$} & \multicolumn{2}{c}{CATE $\tau(C = 1)$} & \multicolumn{2}{c}{CATE $\tau(C=0)$} & \multicolumn{3}{c}{ Reduction in MSE} \\
      &     & Coverage & Power & Coverage & Power & Coverage & Power & $\tau$  & $\tau(1)$   &  $\tau(0)$   \\ \hline

0     & 0.80 & 95\%    & 85\%  & 95\%     & 78\%  & 96\%     & 31\% & 54\%  & 49\%  & 63\% \\
0.1   & 0.78 & 98\%    & 87\%  & 97\%     & 80\%  & 95\%     & 36\% & 58\%  & 53\%  & 63\% \\
0.2   & 0.79 & 94\%    & 89\%  & 94\%     & 80\%  & 95\%     & 41\% & 49\%  & 45\%  & 57\% \\
0.3   & 0.69 & 94\%    & 91\%  & 93\%     & 85\%  & 94\%     & 41\% & 43\%  & 38\%  & 49\% \\
0.4   & 0.47 & 91\%    & 89\%  & 95\%     & 80\%  & 94\%     & 40\% & 26\%  & 28\%  & 39\% \\
0.5   & 0.27 & 92\%    & 83\%  & 94\%     & 76\%  & 95\%     & 32\%  & 14\%  & 16\%  & 34\% \\
0.7   & 0.13 & 94\%    & 73\%  & 94\%     & 65\%  & 94\%     & 27\% & 10\%  & 10\%  & 15\% \\
0.8   & 0.10 & 93\%    & 78\%  & 96\%     & 67\%  & 94\%     & 29\% & 5\%   & 17\%  & 6\%  \\
1     & 0.08 & 93\%    & 72\%  & 95\%     & 61\%  & 94\%     & 26\% & 1\%   & 8\%   & 9\%  \\ \hline
No fusion     &  & 94\%    & 57\%  & 94\%     & 52\%  & 94\%     & 18\% & -  & -  & - \\
\hline
\end{tabular}
\caption{Simulation results for coverage probability of 95\% credible interval, power and reduction in MSE for ATE $\tau$, CATE $\tau(c)$ estimations.}
\label{table:5_2}
\end{table}

\section{PIONEER 6 Application}\label{sec:pioneer6}
We deomonstrate the potential benefit of the proposed method to evaluate the cardiovascular benefits of semaglutide, a novel glucagon-like peptide-1 (GLP-1) analog, in patients with type 2 diabetes at high cardiovascular risk, using the PIONEER 6 trial and a large US health claims dataset.  
PIONEER 6 is a randomized, double-blind, placebo-controlled trial that aims to assess whether oral semaglutide can reduce the incidence of major adverse cardiovascular events (MACE), including heart attack, stroke, and cardiovascular death, compared to placebo. A total of 3,183 participants were enrolled globally across 21 countries between January 2017 and August 2017 and the trial was concluded in September 2018 \citep{pioneer6}. The primary outcome of the PIONEER 6 study was the time from randomization to the first occurrence of MACE. With a hazard ratio of 0.79 (95\% CI, 0.57 to 1.11), the study ruled out an 80\% excess cardiovascular risk with oral semaglutide and confirmed noninferiority to placebo \citep{husain2019oral}.

In this data fusion study, we analyze the occurrence of MACE within one year, targeting the risk difference as the estimand.  
Analyzing the PIONEER 6 data, we found an unadjusted risk difference of $-$1.32\% between treatment and control groups, with a 95\% confidence interval of [$-$2.71\%, 0.06\%], which includes zero. Adjusting for baseline covariates using individual-level trial data improves the precision of the ATE estimate.  \cite{williamson2014variance} showed that inverse probability-of-treatment weighting (IPTW) adjustment always lowers the variance. Therefore, we applied the Augmented Inverse Propensity Weighting (AIPW) \citep{robins1994estimation,rotnitzky1998semiparametric} which involves an outcome model and a propensity model, and we adjusted for key predictors of MACE, including age, HbA1c, high-density lipoprotein (HDL), low-density lipoprotein (LDL), creatinine, albumin, SGLT2i use at baseline, prior revascularization, and heart failure history. This adjustment resulted in an AIPW estimate of the ATE of $-$1.24\% with a narrowed 95\% CI of [$-$2.55\%, 0.06\%], representing a 5.25\% reduction in width. Despite this improvement, as the point estimate moved slightly, the ATE estimate remains statistically non-significant. Given this result and the established non-inferiority of oral semaglutide to placebo, the traditional next step is to proceed with a superiority RCT. Instead, we explore using external real-world data (RWD) to enhance the PIONEER 6 findings.
 
\subsection{Target trial emulation} \label{sec:tte}

In this study, we emulate the PIONEER 6 trial using real-world data sourced from Optum’s linformatics\textsuperscript{\textregistered} Data Mart, consistent with approaches from recent studies  \cite{franklin2021emulating, dang2023case}. The dataset compiles administrative health claims for members with commercial health insurance in the U.S., and includes information on patient enrollment, medical and pharmacy claims, inpatient admissions, and laboratory results. We followed the framework set out in \cite{hernan2016using} and \citep{hernan2022target} to construct a RWD sample that emulates PIONEER 6. We applied the trial’s inclusion criteria to define our RWD sample and established the start of DPP4 inhibitor treatment as ‘time zero’ to accurately mimic trial conditions and control for biases. For covariate adjustment, we employed regularization techniques and Gradient Boosted Machines (GBM) to identify important predictors of treatment, outcome, and for managing the 15.90\% missingness in 1-year MACE outcomes. This emulation process aims to improve the comparability between the RCT and the RWD, thereby increasing the likelihood that Assumption 3 will hold.

\subsection{Design 1: Incorporating control arm only}
We first focus on the design that incorporates only untreated patients from the RWD. We aligned the recruitment window to the emulated RWD dataset with the actual recruitment period of PIONEER 6, followed by selecting records with complete data on key biomarkers (creatinine serum, albumin serum, HbA1c, HDL, and LDL).
Records with covariate values not represented in the PIONEER 6 trial were excluded, leading to a dataset of 3,968 observations. To further enhance comparability, we implemented a final step of 1-to-1 matching, adjusting the covariate distributions to closely match those of PIONEER 6.

The feature selection methods outlined in Section \ref{sec:tte} led to the inclusion of six key variables in our model: age, serum albumin level, serum creatinine level, LDL, HDL and history of heart failure. To address missingness in the outcome, we employed multiple imputation \citep{gelman1995bayesian}. Pooling the posterior samples from five sets of imputations, the ATE estimated from our proposed power likelihood method is $-$1.36\% with a 95\% credible interval of $[-2.42\%,-0.23\%]$. This represents a 20.5\% reduction from the width of the confidence interval from the unadjusted analysis of the PIONEER 6 data. Notably, this enhanced efficiency resulted in the reduction in 1-year MACE risk becoming statistically significant, despite the minimal shift in the point estimate.

\begin{figure}[htbp]
\centering

\begin{minipage}{\textwidth}
    \centering
    \includegraphics[width=0.8\linewidth]{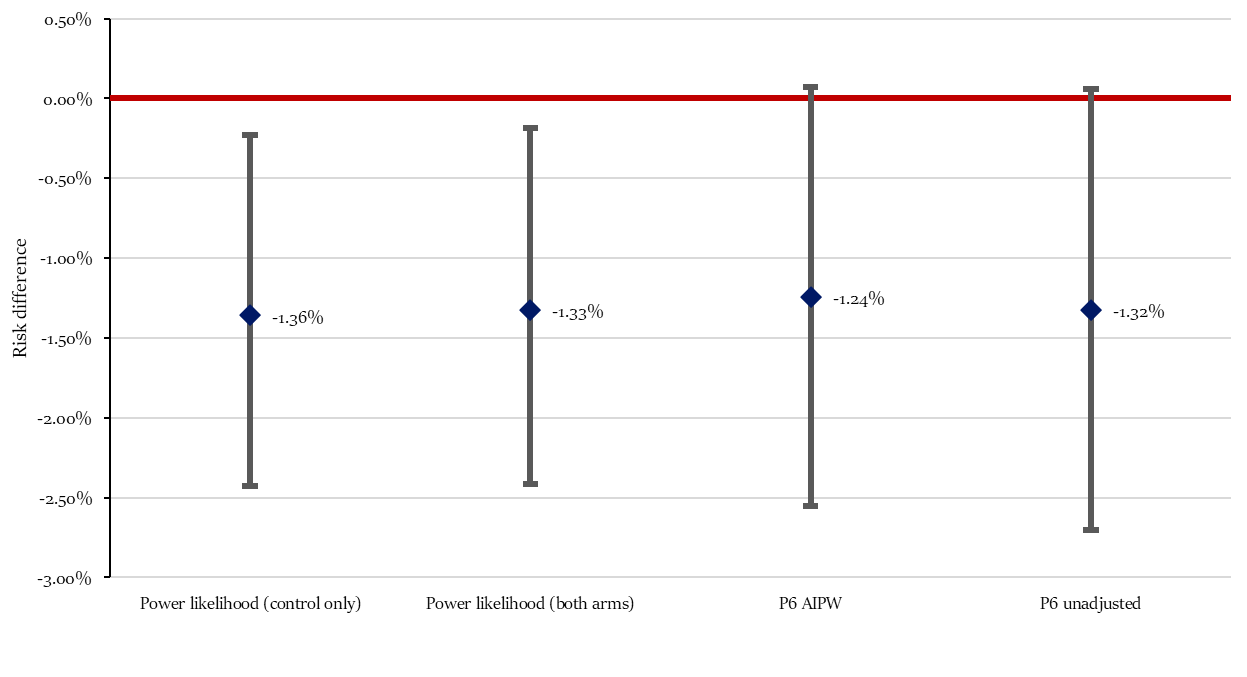} 
\end{minipage}

\vspace{10pt}

\begin{minipage}{\textwidth}
    \centering
    \begin{tabular}{lcc}
    \hline
& Risk difference (\%) & \% of P6 (unadjusted) \\
& (95\% CI) & CI length \\
\hline
Power likelihood (control only) & $-1.36$ $[-2.42, -0.23]$ & 79.54\% \\
Power likelihood (both arms) & $-1.33$ $[-2.41, -0.19]$ & 80.62\% \\
P6 AIPW & $-1.24$ $[-2.55, \phantom{+}0.06]$ & 94.75\% \\
P6 unadjusted & $-1.32$ $[-2.71, \phantom{+}0.06]$ & 100.00\% \\
    \hline
    \end{tabular}
    \caption{Estimated difference in 1-year MACE risk and 95\% confidence/credible interval. P6 unadjusted: Wald test of the MACE rate in the treatment and the placebo arms of PIONEER 6. P6 AIPW: AIPW estimate after adjusting for key covariates in PIONEER 6.}
    \label{table:p6_res}
\end{minipage}

\end{figure}

\subsection{Deisgn 2: Incorporating both arms}
Compared to incorporating only the control arm, including the treatment arm in the RWD posed a unique challenge due to the unavailability of semaglutide in the US until injectable semaglutide was approved by the FDA in December 2017 \citep{FDAOzempic}. Our analysis of the CDM dataset found a substantive uptake only by 2019. This means that aligning the RWD data window with PIONEER 6 recruitment would result in the absence of the treatment. This discrepancy necessitated a broader data window. Moreover, including data from 2019 extended the outcome period into 2020, coinciding with the COVID-19 pandemic outbreak.  A study comprising over 10 million records from the US Department of Veterans Affairs national healthcare databases concluded that COVID-19 infection significantly increases the risk of 1-year MACE (HR 1.55, 95\% CI: [1.50\%, 1.60\%]) \citep{xie2022long}. Considering these factors, we opted for a data window spanning two years before and two years after the PIONEER 6 recruitment period, including outcome data up to September 2020. This led to 89,036 records from the CDM, which met the inclusion and exclusion criteria, of which 2,081 are in the treatment arm. We then applied the same filtering criteria as in the control-only scenario and this left us with 19,206 RWD samples.

Incorporating both arms raised the issue of treatment effect heterogeneity. We aim to neutralize the difference in ATE between the RCT and RWD attributable to the varied distribution of effect modifiers, so that Assumption \ref{assump:transport} is more likelihood to hold and thereby increasing potential efficiency gains from data fusion. Applying the Causal Forest method \citep{wager2018estimation} on the PIONEER 6 data, we identified LDL, insulin use, serum creatinine, and HbA1c as the top variables influencing treatment effect heterogeneity, although the differences were not statistically significant, with the smallest p-value being 0.17, likely due to the limited sample size. Even if the effects were significant, it is crucial to understand that these findings from Causal Forest, such as insulin use, should not be interpreted causally but rather as predictors of treatment effect variation. After including these effect modifiers in the matching process, we obtained a 1:1 matched final RWD sample.

Similarly, we applied the feature selection techniques and ultimately incorporated the same six variables into the model, with an additional one being insulin use. We adjusted for the five effect modifiers in the $P(Y\cmid do(T), 
 \bs C)$ component in the frugal parameterization as outlined in Section \ref{sec:frugal}. With the optimal power $\eta$ adaptively set at 0.6 on average, the estimated ATE is $-1.33\%$ with a 95\% credible interval of $[-2.41\%,-0.19\%]$, a 19.5\% reduction in width from the confidence interval from analysing PIONEER 6 alone. Thus, leveraging augmented RWD, we have concluded a statistically significant reduction in 1-year MACE risk.

 In this case study, as shown in Figure \ref{table:p6_res}, the two data fusion designs yielded consistent results: the ATE's point estimates closely aligned with the mean difference observed in PIONEER 6, yet the confidence intervals effectively shrank by approximately 20\%, demonstrating the effectiveness of our method in robustly augmenting the RCT to increase power.


\section{Discussion}\label{sec:discuss}

Compared to many non-Bayesian data fusion methods, our proposed Bayesian approach offers several advantages 1) adaptivity to alternative causal estimands such as odd ratios, 2) flexibility to incorporate external data with either both arms or only the controls, 3) automatic uncertainty quantification, whic hhas been empirically shown to yield nominal coverage rates approximately. However, a common limitation of Bayesian methods is their difficulty in optimizing the inference of a particular parameter of interest. In contrast, frequentist methods such as targeted learning using TMLE \citep{van2006targeted} and model selection using Focused Information Criteria (FIC) \citep{claeskens2003focused,zhang2011focused}, are designed precisely for this purpose.  In our proposal, we take a typical Bayesian route and select $\eta$ based on expected prediction quality instead of directly targeting the CATE estimates. The loss of efficiency is more pronounced as covariates that are weakly correlated, or not correlated with the outcome get included in the model. Again, this issue is not unique to our method. As mitigation, we recommend employing variable selection techniques toonly keep covariates that are predictive of the treatment and outcomes in the model, as demonstrated in the PIONEER 6 data fusion study.

\bibliographystyle{abbrvnat}
\bibliography{refs}
\appendix


\section{Normal approximation of posterior}\label{sec:app_norm}
\newtheorem{manuallemmainner}{Lemma}
\newenvironment{manuallemma}[1]{%
  \renewcommand\themanuallemmainner{#1}%
  \manuallemmainner
}{\endmanuallemmainner}

\begin{manuallemma}{4.1}



Given the the MLE of $\phi$ in $\mathcal{D}_e$ and $\mathcal{D}_o$ as $\hat \phi_e$ and $\hat \phi_o$ respectively, both are asymptotically normal with $\sqrt{n_e}(\hat\phi_e - \phi^*) \to N(0, V)$ and $\sqrt{n_o}(\hat\phi_o - \phi^* - \delta) \to N(0, W)$, where $\phi^*$ is the true value of $\phi$, $\delta$ represents the `bias' and $V,W \in \mathbb{R}^{m \times m}$.
For $\hat{\phi}_\eta$, the MLE under the $\eta$-powered likelihood,
we find
$
\hat \phi_\eta =  \left(n_eV^{-1} + \eta n_o W^{-1}\right)^{-1}\left(n_eV^{-1}\hat\phi_e + \eta n_o W^{-1}\hat\phi_o \right)
$ and that it has asymptotic variance $V^{-1} + \eta W^{-1}$.
\end{manuallemma}

\begin{proof}
Following the setup outlined in Section~\ref{sec:setup}, we have an experimental dataset of size $n_e$, $X = (X_1,\dots,X_{n_e})$ and an observational dataset of size $n_o$, $Y = (Y_1,\dots,Y_{n_o})$. We assume that they share the same causal parameters $\phi^*$.

We consider the following estimation equations:
\begin{align}
    \sum_{i=1}^{n_e} f(\phi; X_i) &= 0 , \label{eqn:ee1}\\
    \sum_{i=1}^{n_o} g(\phi; Y_i) &= 0 ,\label{eqn:ee2}\\
    \sum_{i=1}^{n_e} f(\phi; X_i) + \eta \sum_{i=1}^{n_o} g(\phi; Y_i) &= 0 \label{eqn:ee3},
\end{align}
where $\E f(\phi^*; X_i) = 0$ and $\E g(\phi^* + \delta; Y_i) = 0$ for some unknown bias $\delta$. From the standard theory of estimation equations, we know that the solution to (\ref{eqn:ee1}), $\hat\phi_e $, is asymptotically normal following $\sqrt{n_e}(\hat\phi_e - \phi^*) \to N(0, V)$, and likewise, $\sqrt{n_o}(\hat\phi_o - \phi^* - \delta) \to N(0, W)$.

Let the solution to (\ref{eqn:ee3}) be $\hat \phi_\eta$. We can approximate each sum in (\ref{eqn:ee3}) with its Taylor expansion and use the fact that $\left [ \sum_{i=1}^{N_e} f(\hat\phi_e; X_i) \right] = 0$ and $\left [ \sum_{i=1}^{N_o} g(\hat\phi_o; X_i) \right] = 0$, then 
\begin{align*}
    n_e V^{-1} (\hat \phi_\eta - \hat\phi_e) + \eta n_o W^{-1} (\hat \phi_\eta - \hat\phi_o) &= 0\\
    \left(n_eV^{-1} + \eta n_o W^{-1}\right) \hat \phi_\eta &= n_eV^{-1}\hat\phi_e + \eta n_o W^{-1}\hat\phi_o.
\end{align*}
Then we have an expression for $\hat \phi_\eta$ as
$$
\hat \phi_\eta =  \left(n_eV^{-1} + \eta n_o W^{-1}\right)^{-1}\left(n_eV^{-1}\hat\phi_e + \eta n_o W^{-1}\hat\phi_o \right).
$$
Let $\mathcal{I}(\hat \phi_e)$ and  $\mathcal{I}(\hat \phi_o)$ be the Fisher information matrices (FIMs) for the experimental and observational data respectively, then the combined FIM is
\begin{align*}
    \mathcal{I}(\hat \phi_\eta) &= -\mathbb{E}\left[\frac{\partial}{\partial \phi}\left (\sum_{i=1}^{n_e} f(\phi; X_i) + \eta \sum_{i=1}^{n_o} g(\phi; Y_i) \right)\right]\\
    &= -\mathbb{E}\left[\frac{\partial}{\partial \phi} \sum_{i=1}^{n_e} f(\phi; X_i)\right]  -\eta \, \mathbb{E}\left[\frac{\partial}{\partial \phi}\sum_{i=1}^{n_o} g(\phi; Y_i)\right]\\
    &= \mathcal{I}(\hat \phi_e) + \eta \, \mathcal{I}(\hat \phi_o).
\end{align*}
That is, the combined Fisher Information is a weighted sum of the two datasets.

\end{proof}

\section{Proof of Theorem 1}\label{sec:app_proof}
\newtheorem{manualtheoreminner}{Theorem}
\newenvironment{manualtheorem}[1]{%
  \renewcommand\themanualtheoreminner{#1}%
  \manualtheoreminner
}{\endmanualtheoreminner}

\begin{manualtheorem}{1}[\textbf{Consistency}]

Let $\delta = \delta(n_o) = \delta^*/n_o^k$ denote the bias in $\D_o$.
%
\begin{enumerate}[label = (\alph*)]
    \item If $0 \leq k < 1/2$ and $\delta^* \neq 0$, then $\hat \eta$ converges to 0, and $\hat \phi_{\hat \eta} = \phi^*+ o_p(n^{-1/2})$. 
    \item If $k \geq 1/2$ or $\delta^* = 0$ then $\hat \eta $ does not converge.

\end{enumerate}
\end{manualtheorem}

\begin{proof}
From \citet{dunsmore1976asymptotic}, we have that the posterior predictive distribution can be Taylor expanded to:
\begin{align*}
    p(\tilde x \cmid  x) = p(\tilde x \cmid \hat\phi) + \frac{1}{2n} I_p(\hat\phi)^{-1} \frac{\partial^2 p(x \cmid \hat{\phi})}{\partial \phi^2} + O_p(n^{-3/2}),
\end{align*}
where we note that this does not depend upon the prior distribution that we choose.  

As discussed in Remark \ref{rmk:elpd}, ELPD is a proper scoring rule, so it will maximize the expected score at the distribution that is closest to the truth. That is, as sample size tends to infinity, the value of $\phi$ that maximizes ELPD converges to $\phi^*$.

From Lemma \ref{lem:norm_approx} we have that 
\begin{align*}
    \hat \phi_\eta &=  \left(n_eV^{-1} + \eta n_o W^{-1}\right)^{-1}\left(n_eV^{-1}\hat\phi_e + \eta n_o W^{-1}\hat\phi_o \right)\\
    & = \left(n_eV^{-1} + \eta n_o W^{-1}\right)^{-1}\left \{n_eV^{-1}\left(\phi^* + o_p(n^{-{1/2}}) \right) + \eta n_o W^{-1}\left(\phi^* +\delta^* + o_p(n^{-{1/2}}) \right) \right \}\\
    & = \phi^* + \left(V^{-1} + \frac{\eta}{\rho} W^{-1} \right) \frac{\eta}{\rho} \delta^* n^{-k} + o_p(n^{-1/2}). 
\end{align*}
In scenario (a), $k < \frac{1}{2}$ and we have that 
\begin{align*}
    \hat \phi_\eta &= \phi^* + \left(V^{-1} + \frac{\eta}{\rho} W^{-1} \right) \frac{\eta}{\rho} \delta^* n^{-k} + o_p(n^{-1/2}).
\end{align*}
Then, since ELPD is a proper scoring rule, and the bias is 
statistically non-negligible, the optimal $\eta$ will asymptotically
be zero.  

In scenario (b), $k \geq \frac{1}{2}$ or $\delta^* = 0$ and we have that 
\begin{align*}
    \hat\phi_\eta &= \phi^* + \eta \delta^* n^{-k} + o_p( n^{-1/2})  = \phi^* + o_p( n^{-1/2}), 
\end{align*}
and $\eta$ has no asymptotically and statistically relevant effect on the 
ELPD at all.  Hence we would not expect $\eta$ to converge.
\end{proof}



\end{document}